\documentclass[
reprint,
amsmath,amssymb,
aps,
]{revtex4-2}
\usepackage{CJK}
\usepackage{graphicx}
\usepackage{dcolumn}
\usepackage{bm}
\usepackage{makecell}
\usepackage{amsmath}
\usepackage{hyperref}
\usepackage{color}
\usepackage{multirow}
\usepackage{array}
\usepackage{booktabs}
\usepackage{upgreek}
\usepackage{physics}

\begin{document}
\begin{CJK*}{UTF8}{gbsn}	
	\preprint{APS/123-QED}
	\title{Closed-loop dual-atom-interferometer inertial sensor \\ with continuous cold atomic beams}
 
	\author{Zhi-Xin Meng({\CJKfamily{gbsn}孟至欣})$^{1,2,4}$, Pei-Qiang Yan({\CJKfamily{gbsn}颜培强})$^{1,2}$, Sheng-Zhe Wang({\CJKfamily{gbsn}王圣哲})$^{1,2}$, Xiao-Jie Li({\CJKfamily{gbsn}李晓杰})$^{1,2}$, Hong-Bo Xue({\CJKfamily{gbsn}薛洪波})$^{3}$, Yan-Ying Feng({\CJKfamily{gbsn}冯焱颖})$^{1,2}$
    }
    \email{yyfeng@tsinghua.edu.cn}
    \affiliation{
    $^{1}$State Key Laboratory of Precision Measurement Technology and Instruments, Tsinghua University, Beijing 100084, China
    }%
    \affiliation{
    $^{2}$Department of Precision Instrument, Tsinghua University, Beijing, 100084, China
    }
    \affiliation{
    $^{3}$State Key Laboratory of Space Weather, National Space Science Center, Chinese Academy of Sciences, Beijing 100190, China
    }%
	\affiliation{
    $^{4}$Beijing Institute of Aerospace Control Devices, Beijing 100039, China
    }%
    
	\date{\today}
	
	\begin{abstract}
		We demonstrate a closed-loop light-pulse atom interferometer inertial sensor that can realize continuous decoupled measurements of acceleration and rotation rate. The sensor operates with double-loop atom interferometers, which share the same Raman light pulses in a spatially separated Mach-Zehnder configuration and use continuous cold atomic beams propagating in opposite directions from two 2D$^+$ magneto-optical trappings. Acceleration and the rotation rate are decoupled and simultaneously measured by the sum and difference of dual atom-interferometer signals, respectively. The sensitivities of inertial measurements are also increased to be approximately 1.86 times higher than that of a single atom interferometer. The acceleration phase shift is compensated in real time by phase-locking these interferometers via the Raman laser phases from the sum interferometer signal, and the gyroscope perfomance is improved.  We achieve long-term stabilities of $6.1 \ \mu g$ and 840 nrad/s for the acceleration and the rotation rate, respectively, using a short interrogation time of 0.87ms (interference area $A=0.097$ mm$^2$). This work provides a building block for an atomic interferometer based inertial measurement unit for use in field applications that require a high data-rate and high stability.

	\end{abstract}
	

	\maketitle
    \end{CJK*}
	\section{Introduction}
	Over more than 30 years of development \cite{kasevich1991atomic, riehle1991optical, peters2001high, durfee2006long}, light-pulse atom interferometers (LPAIs) have demonstrated their potential for use as high-sensitivity quantum sensors for measuring inertial quantities, including acceleration\cite{mcguinness2012high, lautier2014hybridizing}, rotation rate \cite{durfee2006long, savoie2018interleaved}, gravity, and gravity gradient \cite{niebauer1995new, peters2001high, sorrentino2012simultaneous}. Various applications are available for LPAI-based inertial sensors, such as high-precision inertial navigation \cite{gustavson1997precision, canuel2006six, stockton2011absolute}, fundamental physics \cite{weiss1993precision, yu2019atom, asenbaum2020atom, di2021gravitational, arvanitaki2018search}, geophysics \cite{andersen2005global, jacob2008absolute}, and quantum metrology \cite{rosi2014precision}. In recent years, attempts have been made to move LPAI-based inertial sensors from laboratory environments to field applications \cite{carraz2009compact, bidel2018absolute, bongs2019taking, lachmann2021ultracold}, but several bottleneck challenges have yet to be overcome, e.g., problems with low data rates and low dynamic ranges \cite{rakholia2014dual, lautier2014hybridizing, bonnin2018new, avinadav2020composite, yankelev2020atom}.
	
	After the early stage of development, in which thermal atomic beams were used as matter-wave sources \cite{kasevich1991atomic,durfee2006long}, most research into atom interferometry focused on using of a pulsed cold atom source \cite{geiger2020high}. For inertial sensing, these cold LPAIs offer advantages in terms of interrogation time and system volume over LPAIs with thermal-atom beam sources because they have slower longitudinal mean velocities and narrower longitudinal velocity distributions \cite{yver2003reaching,dutta2016continuous}. Furthermore, the narrower horizontal velocity distribution of a cold atomic cloud usually produces a higher static fringe contrast because more atoms are involved in the Doppler-sensitive Raman transition and thus contribute to the interferometer signal \cite{muller2007versatile, gauguet2009characterization}. A cold LPAI gyroscope has been demonstrated with sensitivity and stability that can compete with those of the best strategic-grade fiber-optic gyroscopes \cite{savoie2018interleaved}. However, it is a major challenge for LPAIs with pulsed cold atomic sources to measure time-varying signals because of their low data rates ($\sim$ a few Hz) \cite{gauguet2009characterization,muller2009compact,savoie2018interleaved}, which suffer from their time sequential operating mode and the loading times required for the magneto-optical traps (MOTs). Low data rates and low bandwidths are bottleneck problems that limit the application of LPAI-based inertial sensors in dynamic environments, especially application to high-precision inertial navigation.
	
	Among the various trials of high-date-rate LPAIs \cite{rakholia2014dual,lautier2014hybridizing}, LPAIs based on continuous atomic beams have shown the potential to achieve higher data rates ($\sim$ 100 Hz) without loss of sensitivity \cite{durfee2006long}. A high data rate makes it easier for an LPAI inertial sensor to tune the interferometer's phase using Raman frequencies or phases. A continuous operating mode is also helpful for performing real-time compensation of the rotation rate and acceleration and realizing a closed-loop system, which generally demonstrates a broader dynamic range, better robustness, and improved long-term stability when compared with open-loop systems. In addition, LPAIs with continuous atomic beams inherently work in a zero-dead-time operating mode that cancels the aliasing noises caused by the Dick effect \cite{joyet2012theoretical}. The first continuous cold LPAI was demonstrated by our group using a low-velocity intense source (LVIS) of cold atoms \cite{xue2015continuous}. An LPAI with a continuous three-dimensional (3D) cold atomic beam operating at sub-Doppler temperatures (15 $\mu$K) was demonstrated recently with an inferred short-term phase measurement noise of 530(20) $\rm{{\mu rad}/{\sqrt{Hz}}}$ \cite{kwolek2020three,kwolek2022continuous}.
	
	Among the different measurement protocols, the fringe fitting technique leads to great long-term stability for an atom-interferometer gyroscope \cite{durfee2006long,merlet2009operating,tackmann2014large}. When compared with fringe fitting technique, the closed-loop approach offers better time resolution to achieve a shorter time constant for the lock loop \cite{merlet2009operating}. A closed-loop system locked into mid-fringe operation has been realized on an atom-interferometer gyroscope with two atomic clouds \cite{tackmann2014large}. By feeding the error signal back to the common piezo stepper motor mounted on the Raman mirrors, the acceleration phase of the second interferometer can be corrected for slow drift. Another closed-loop technique uses the mid-fringe locking to cause LPAIs to work within a linear regime with maximal sensitivity \cite{deng2017common,savoie2018interleaved,yao2021self}. This approach was realized by modulating the interferometer phase by $\pm \pi/2$ and feeding the error signal back onto the Raman relative phase, which is computed from two successive alternate measurements taken on both sides of a fringe. These locking schemes demonstrate advantages that include shorter time constants, higher sensitivity, improved robustness, and long-term stability \cite{merlet2009operating,deng2017common}. Closed-loop locking technology is easier to implement in LPAIs with continuous atomic beams because of the successive output signals.
	
	In this work, we report the experimental realization of an LPAI inertial sensor with dual closed-loop phase-locked interferometers. Simultaneous, decoupled, and continuous measurements of the accelerations and rotation rates are realized by the sensor using continuous cold atomic beams of $^{87}$Rb generated from 2D$^+$ MOTs and Raman laser beams with Mach-Zehnder-type geometry. The differential dual-atom-interferometer operating mode effectively suppresses the common-mode noise of the gyroscope, and improves the measurement sensitivity. The vibration noise of the gyroscope is suppressed by feeding the sum signal back in a closed-loop mode. We measure the acceleration and rotation rate, and evaluate the sensor's long-term stabilities in terms of Allan deviations. Long-term stabilities of $6.1 \ \mu g$ for the acceleration and 840 nrad/s for the rotation rate are demonstrated with a spatially separated interference length of only $L=9.5$ mm and a corresponding interrogation time of $T=0.87$ ms. Furthermore, the LPAI inertial sensor can track a continuously changing rotation rate well by applying an external force on the platform, thus enabling real-world sensor applications.
	
	\section{Experimental setup}
	\begin{figure*}[]
		\includegraphics[width=0.8\linewidth]{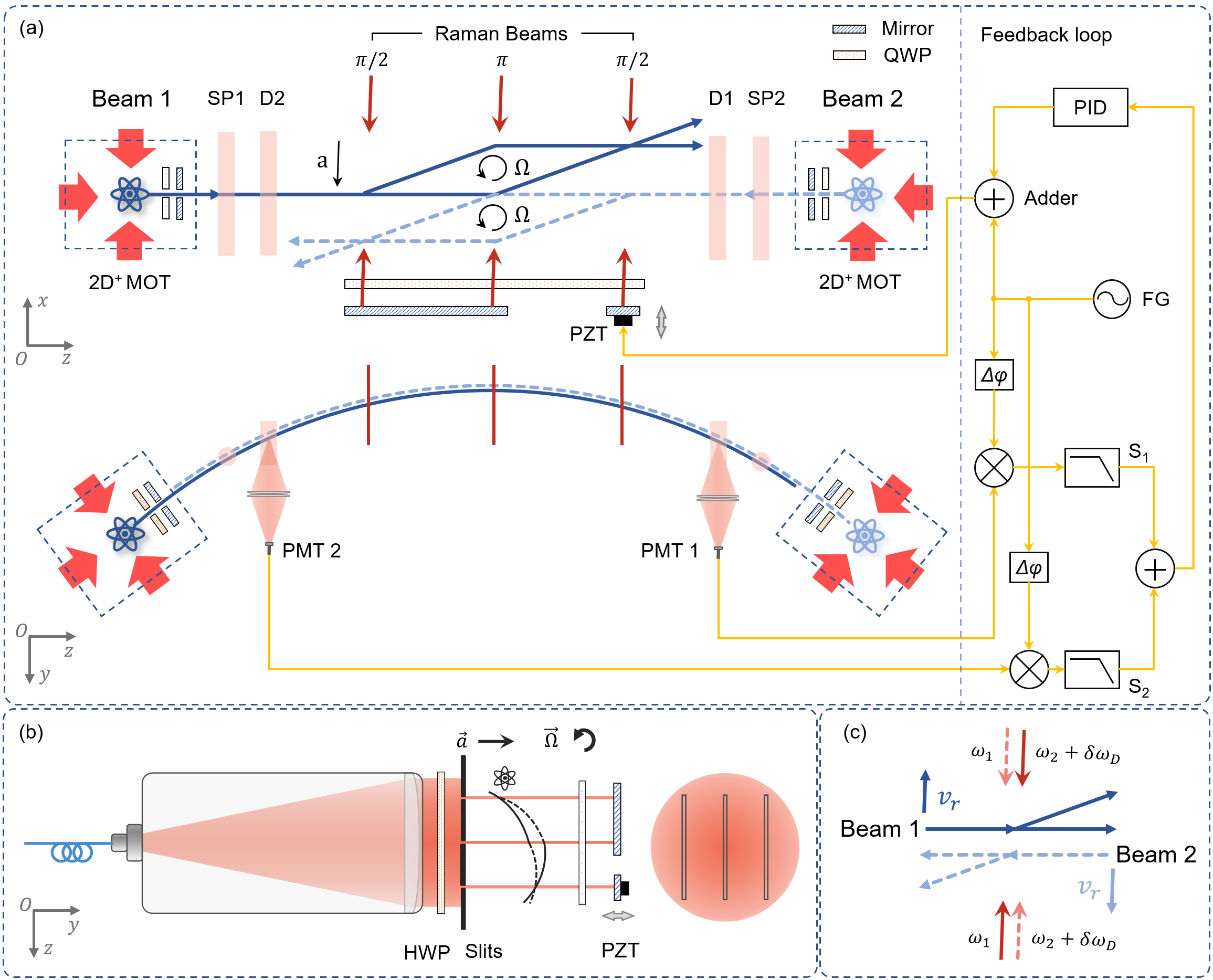}
		\caption{\label{fig:whole system} (a) Closed-loop dual-AI inertial sensor based on two continuous cold atomic beam sources, which are generated from two 2D$^+$ MOTs and are ejected along the same parabolic trajectory but in opposite directions. Three spatially separated Raman light pulses with a $\frac{\pi}{2}-\pi-\frac{\pi}{2}$ sequence coherently split, reflect, and recombine the atomic wavepackets for two atomic beams after state preparation(SP1 or SP2), and form two spatial domain AIs of the Mach-Zehnder type. The interference phase shifts are measured using two photomultipliers (PMT1 or PMT2) by collecting the fluorescence induced by the detection light (D1 or D2). Two interference signals from two counterpropagating atomic beams are used to perform differential detection and to eliminate the influence of common mode noise. The right side of (a) shows the feedback loop of the AI inertial sensor. We use a modulation method to increase the signal-to-noise ratio (SNR) by adding a sinusoidal modulation at 81 Hz to the PZT. The added demodulated output signals of the dual AIs are then fed back to the PZT driver via a PID controller to track the interference phase to the set point in real time. (b) Three spatially separated Raman pulses are formed using a mask with three slits (dimensions of 1 mm $\times$ 20 mm) after a specifically designed fiber collimator. The corresponding interference area is 0.097 mm$^2$. The spacing between the three slits is 9.5 mm, which leads to a Raman laser intensity ratio of 1:2:1. All three counterpropagating Raman beams are parallel to within approximately 80 $\mu$rad. (c) Because the detuning of the Raman laser is set to compensate for the Doppler frequency shift, each of the two counterpropagating atomic beams can only interact with only one pair of beams and thus acquire the recoil momentum in opposite directions. QWP, Quarter wave plate. HWP, Half wave plate. The image here is not to scale.}
	\end{figure*}

   	The configuration of the proposed closed-loop dual-atom-interferometer inertial sensor is shown in Fig. \ref{fig:whole system}. Two continuous cold $^{87}$Rb atomic beams, designated 1 and 2, are generated from 2D$^+$ MOTs located at both ends of the sensor. These beams propagate along the same path but in opposite directions, as demonstrated in our previous work \cite{chen2017tuning,meng2020atom,wang2023atom}. The vacuum apparatus volume is approximately 1.5 m$\times$0.3 m$\times$0.3 m, and two ion pumps are used to evacuate the detection and interferometry zone. In the 2D$^+$ MOT zone, a set of four rectangular coils generates a quadrupole field with a line of zero magnetic field along the symmetry axis. Two pairs of counterpropagating cooling laser beams are perpendicular to each other with a power of 200 mW and size of 100 mm$\times$25 mm. A continuous cold atomic beam is generated through an optical pressure imbalance, that is induced by a 30-mW pushing laser beam ($\phi$25 mm) and a retro-refecting mirror with a 1-mm hole drilled at its center.
    
    The fluxes of these cold atomic beams are measured to be up to $2\times{10}^9$ atoms/s with a mean longitudinal velocity of 10.9(1) m/s and a longitudinal velocity distribution of 3.0(1) m/s using the time-of-flight (TOF) method. We measured Doppler-sensitive Raman transition spectra of cold atomic beams using a counterpropagating Raman beam with a 1mm width and a detection light situated approximately 600mm from the exit of the 2D$^+$ MOT. The linewidth (full width at half maximum, FWHM) was calculated to be $2\pi \cdot (224\pm4)$ kHz, corresponding to an effective atomic transverse temperature of 14.1(5) $\mu$K. The linearly polarized state preparation laser beams prepare approximately 85\% of the atoms into the $\left|F=1, m_{F}=0\right\rangle$ ground state with wavevectors oriented parallel to the direction of the Raman bias magnetic field, tuned to the $\left| F=2\right\rangle \rightarrow\left|F^{\prime}=1\right\rangle$ and $\left|F=1\right\rangle \rightarrow \left|F^{\prime}=0\right\rangle$ D$_2$ transition line.
    
    Three spatially separated Raman light pulses with a $\frac{\pi}{2}-\pi-\frac{\pi}{2}$ sequence coherently split, reflect, and recombine the atomic wave-packets for two atomic beams by stimulated Raman transitions, and form two spatial domain atom interferometers (AIs) of the Mach-Zehnder type. The interference phase shifts are measured via the atomic numbers populated in the $\left|5^{2} S_{1/2}, F=2\right\rangle$ state by collecting light-induced fluorescence signals using PMTs (H7422-50, Hamamatsu, Japan) with a detection laser beam that has a power of 250$\mu$W and size of 20 mm$\times$4 mm. Each of the two AIs operates as an inertial sensor that is sensitive to the acceleration $\mathbf{a}$ along the direction of the Raman effective wavevector $\mathbf{k}_{\mathbf{eff}}$ and the rotation $\mathbf{\Omega}$ along the normal direction of the interference area. The AI phase shifts are modulated at a frequency of 81 Hz using a piezoelectric transducer (PZT; PAL 20 VS12, NanoMotions, China), mounted behind one of the planar mirrors for the $\pi/2$ Raman pulse. The other two Raman laser beams are reflected by a rectangular mirror mounted on a home-made mount. The demodulated signals acquired via two lock-in amplifiers (LIAs) from the two AIs are added and then fed to a proportional-integral-derivative (PID) controller. Then the signal from the PID controller is fed back to the PZT driver for real-time phase-shift compensation. The maximum measurement sensitivity is realized by locking the zero crossing of the interference signals\cite{ebberg1985closed,pavlath1996closed,yao2021self}.
    
	The Raman beam is detuned by $-$1.23 GHz from the master laser frequency using a double-pass acousto-optic modulator (AOM; GPF-650, Brimrose, USA). The two Raman-beam frequencies are generated using a fiber electro-optic phase modulator (fEOM; PM-0K5-10, EOSPACE, USA). After power amplification by a tapered amplifier (TA; BoosTA, Toptica, Germany), the Raman beams are shaped into a Gaussian beam with a diameter of 50 mm using a specifically designed fiber collimator, as shown in Fig. \ref{fig:whole system}(b). In our experiment, three spatially separated Raman pulses are formed that have a laser intensity ratio of 1:2:1, a beam size of 1 mm $\times$ 20 mm, and a spacing of 9.5 mm, using a mask with three slits fabricated in parallel in a single substrate \cite{xue2015continuous}. The corresponding interference area is 0.097 mm$^2$. The spatial parallelism of the three Raman beams is important for realizing the Doppler-sensitive AI and obtaining high contrast \cite{kasevich1991atomic2,kwolek2022continuous}. In our case, the spatial parallelism is theoretically determined to be approximately 25 $\mu$rad based on the far-field divergence angle of the laser output from the collimator and optimized experimentally to be less than 30 $\mu$rad with a shearing interferometer (SI500, Thorlabs, USA). The alignment and the parallelism of the retro-reflected Raman beams are finely adjusted using the mirrors by monitoring the output power at the entrance of a fiber-optic beam splitter that is reversely inserted between the TA and the collimator. The parallelism of the counterpropagating Raman beams is estimated to be approximately 25 $\mu$rad based on the mode field diameter of the optical fiber and the optical path geometry. All three counterpropagating Raman beams are parallel to within 80 $\mu$rad.
    
	The Raman beams generated by the fEOM allow one atomic beam to experience two diffraction processes that have Raman effective wavevectors with opposite signs. To select one specific trajectory, the Raman beams are tilted by 0.6$^{\circ}$ from the direction perpendicular to the longitudinal mean velocity of the atoms in the $Oxz$ plane, and the two-photon detuning process is set to compensate for the Doppler frequency shift. By sharing the same Raman pulse sequence, two AIs can form with opposite Raman effective vectors $\mathbf{k}_{\mathbf{eff}}$ and $-\mathbf{k}_{\mathbf{eff}}$, because of the different Doppler detunings for the two counterpropagating atomic beams, as shown in Fig. \ref{fig:whole system}(c). Both AIs are sensitive to the rotation rates oriented perpendicular to the paper along the same direction as the interference areas and the linear accelerations along the directions of the Raman wavevectors. The output interference signals $S_{1}$ and $S_{2}$ from the two interferometers are related to the linear acceleration $\mathbf{a}$, the rotation rate $\mathbf{\Omega}$, the non-inertial initial phase shift $\phi_0$, and the non-inertial phase shift $\phi_r$ in the reflected path:	
	\begin{equation}
		\begin{aligned}
			&S_{1}=A_{1}+B_{1} \cos \left(\phi_{a}-\phi_{\Omega}-\phi_{r}+\phi_{0}\right)+n(t), \\
			&S_{2}=A_{2}+B_{2} \cos \left(-\phi_{a}-\phi_{\Omega}+\phi_{r}+\phi_{0}\right)+n(t).
		\end{aligned}
	    \label{equ:basic}
	\end{equation}
	where $A_i$ and $B_i$ (i=1,2) are the offset and the amplitude of interference signal $S_i$ (i=1,2), respectively. $\phi_\mathrm{\Omega}=2\mathbf{k}_{\mathbf{eff}} \cdot \mathbf{\Omega}\times\mathbf{v}T^2$ is the Sagnac phase shift caused by the rotation rate $\mathbf{\Omega}$ and $\phi_a=\mathbf{k}_{\mathbf{eff}} \cdot \mathbf{a}T^2$ is the phase shift caused by the linear acceleration $\mathbf{a}$. The vector $\mathbf{v}$ is the velocity of the atoms, and $T$ is the interrogation time. $n(t)$ is the common-mode noise caused by detection, the signal acquisition system, and other sources. For the dual AIs, $\phi_a$ and $\phi_r$ have opposite signs because of their different Raman effective vectors, whereas they have the same rotation phase shift $\phi_\Omega$ because the directions of the atomic velocity and the Raman effective vectors are opposite in both cases. The dual AIs also have the same non-inertial initial phase shift $\phi_0$.
	
	When the Raman mirror is scanned with the PZT in one direction, the Raman laser phases are tuned in the following form:
	\begin{equation}
		\phi_{r}=2 \pi \epsilon t+\phi_{r}^{0},
	\end{equation}
	where $\epsilon$ is the scanning rate of the Raman laser phase, $\phi_r^0$ is the initial phase shift of the reflection loop, which is related to the initial positions of the mirrors. The difference between $A_1$ and $A_2$ and that between $B_1$ and $B_2$ come mainly from inconsistencies in the LIA parameters, the PMT magnifications, and the atomic fluxes between the two AIs. By setting the parameters of the PMTs and LIAs, such that $A_1=A_2=A$ and $ B_1=B_2=B$, we can write the sum and differential signals as:
	\begin{equation}
		\begin{aligned}
			S_{1}+S_{2}=&
			2 A+2 B \cos \left(\phi_{a}-2 \pi \epsilon t-\phi_{r}^{0}\right) \cos \left(\phi_{\Omega}-\phi_{0}\right)\\
			&+2 n(t), \\
			S_{1}-S_{2}=&
			-2 B \sin \left(\phi_{a}-2 \pi \epsilon t-\phi_{r}^{0}\right) \sin \left(\phi_{\Omega}-\phi_{0}\right).
		\end{aligned}
		\label{equ:sum and add}
	\end{equation}
	
	\begin{figure}[]
		\includegraphics[width=\linewidth]{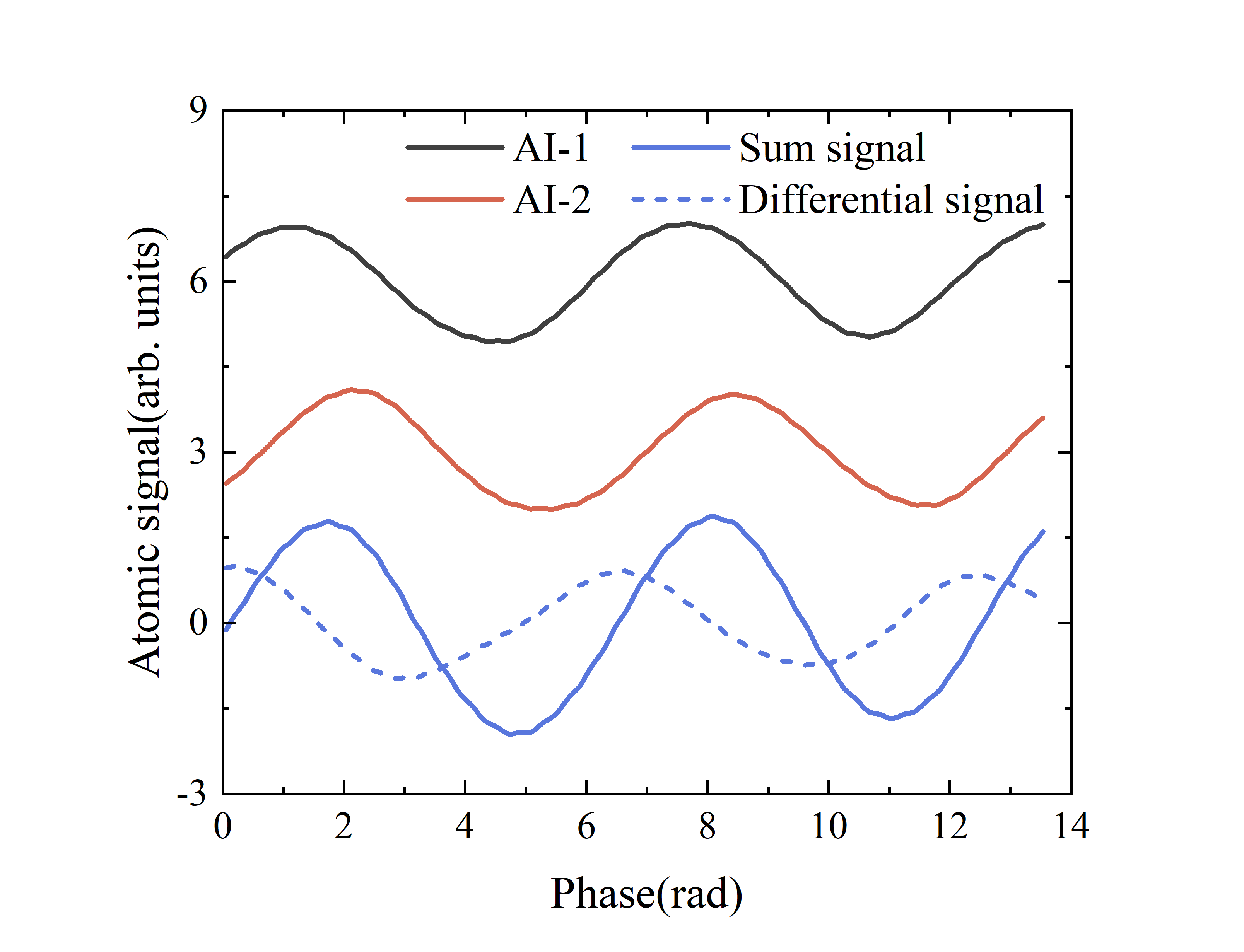}
		\caption{\label{fig:two fringes} Raman-MZ AI fringes. The optical phase of one of the $\frac{\pi}{2}$ Raman pulses is scanned to obtain the dual AI fringes (black and red). The blue solid line and dashed line are the sum signal and the differential signal, respectively.}
	\end{figure}

	Figure \ref{fig:two fringes} shows the typical interference fringes obtained with an interrogation time of $T=0.87$ ms when scanning and modulating the relative Raman laser phase of one of the $\frac{\pi}{2}$ Raman pulses. The amplitudes of the sum signal (blue solid line) and the differential signal (blue dashed line) of the two AIs depend on the rotation phase shift $\phi_\Omega$, the initial phase shift $\phi_0$ of the Raman laser, and the amplitude $B$ of the two interference signals, according to Eq. (\ref{equ:sum and add}). The amplitudes of the sum signal of both AIs is $\sim$1.86 times that of a single AI signal in this type of measurement, which leads to higher inertial sensitivity. As the initial phase shift $\phi_0$ of the Raman laser is around 0, the amplitudes of the differential signal is smaller. In addition, there is a constant phase difference of $\frac{\pi}{2}$ between the differential and sum signals, which means that the zero point of the differential signal corresponds to the maximum of the sum signal.
	
	When the interferometer phase shift of the inertial sensor is locked with a phase control signal $\Delta\phi$, which is fed back to the atomic interference phase, Eq. (\ref{equ:sum and add}) becomes:
	\begin{equation}
		\begin{aligned}
			S_{1}+S_{2}=&
			2 A+2 B \cos \left(\phi_{a}-\Delta\phi-\phi_{r}^{0}\right) \cos \left(\phi_{\Omega}-\phi_{0}\right)\\
			&+2 n(t), \\
			S_{1}-S_{2}=&
			-2 B \sin \left(\phi_{a}-\Delta\phi-\phi_{r}^{0}\right) \sin \left(\phi_{\Omega}-\phi_{0}\right).
		\end{aligned}
		\label{equ:sum and different singal}
	\end{equation}
	
	According to Eq. (\ref{equ:sum and different singal}), if the sum signal of the dual AIs, i.e., $S_1+S_2$, is fed back for the closed-loop measurement, changes in the acceleration will cause the PID controller to output additional compensation to maintain the error signal at zero. Then the Raman laser phase shift controlled by the PZT in the reflected path will compensate for the acceleration phase shift in real time, and the acceleration can subsequently be calculated using the output signal from the PID controller. Because the error signal will always be locked at the zero point, the differential signal leads to a phase shift of $\phi_a-\Delta\phi-\phi_{r}^{0}=0$. Additionally, the differential signal, $S_1-S_2$, gives the rotation rate during the real-time acceleration compensation, and the differential detection mode suppresses the common-mode noise. This method realizes decoupled measurements of the rotation rate and the acceleration using dual AIs with counterpropagating atomic beam sources.
	
	\section{Results and Discussion}
\begin{figure}[]
		\includegraphics[width=\linewidth]{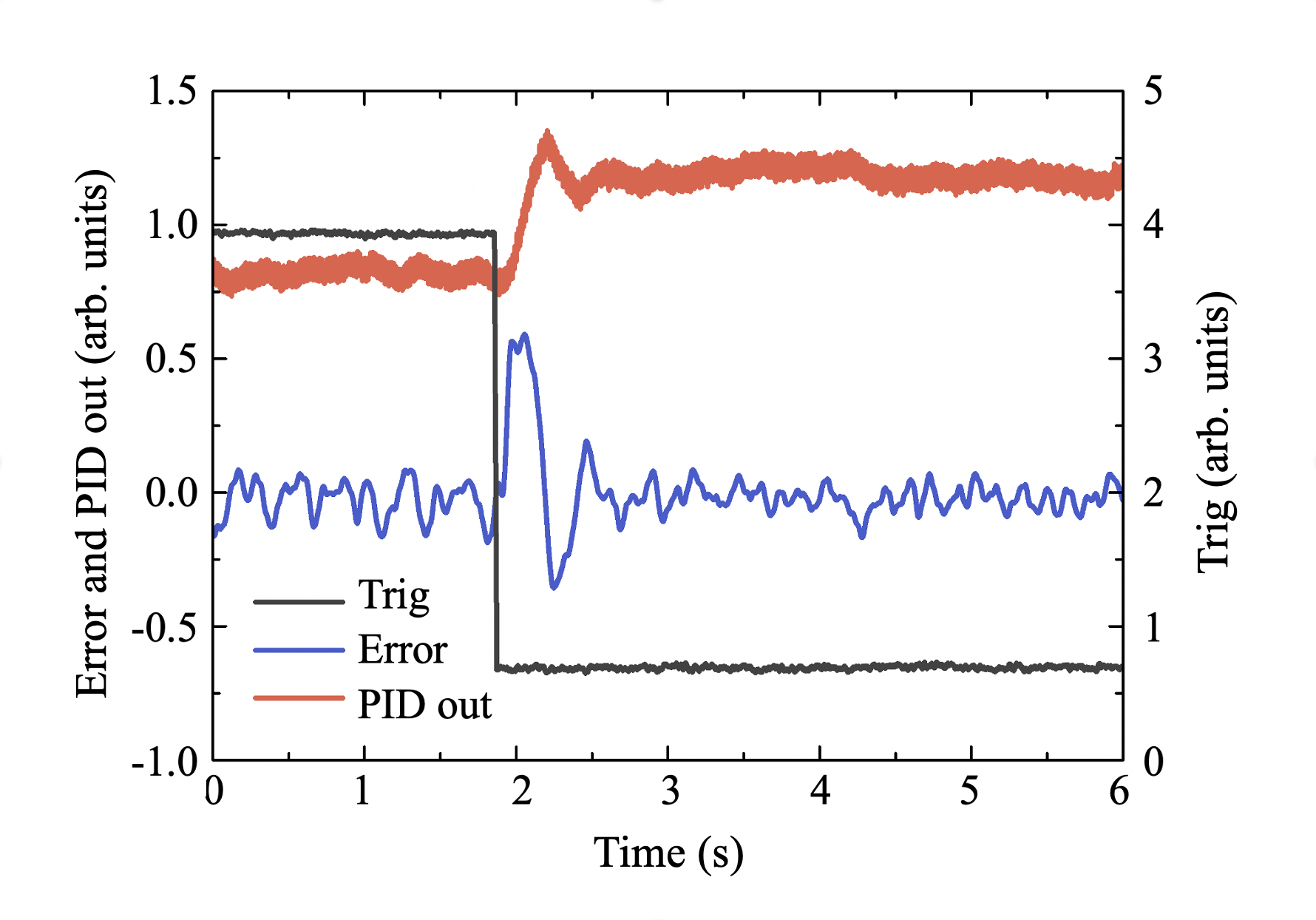}
		\caption{\label{fig:dynamic response} Dynamic response of the closed-loop AI when the direction of each Raman effective wavevector is switched. After a phase jump, the PID controller responds over time, and the AI inertial sensor continues to work near the set point.}
	\end{figure}

	To verify the effectiveness of the closed-loop control system, an excitation was added to the laser phase. Figure \ref{fig:dynamic response} shows a typical step response, in which an abrupt change in the inertial phase shift is caused by switching the directions of the Raman effective wavevectors $\pm\mathbf{k}_{\mathbf{eff}}$. The PID controller output (red line) produces an effective response to compensate for the abrupt change in the phase shift, as also demonstrated by the error signal (blue line). The closed-loop system tracks the transient input and then enters a new stable state in approximately 1 s, which makes it possible to compensate the vibration noise at frequencies below 1 Hz. The real-time vibration-noise compensation gives the gyroscope a stronger anti-jamming ability and increases robustness in a dynamic environment. 

        \begin{figure}[t]
		\includegraphics[width=\linewidth]{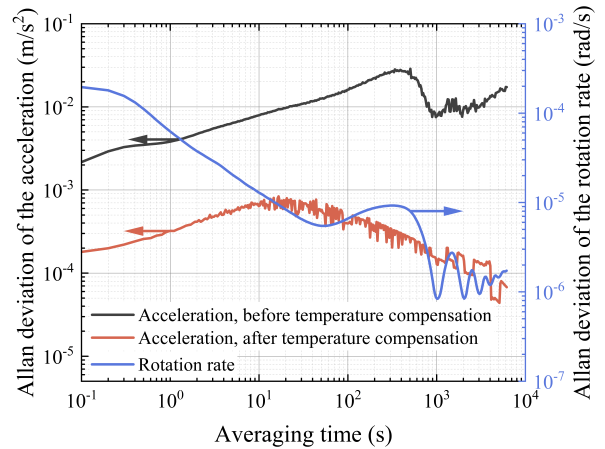}
		\caption{\label{fig:rotation rate} Performance characteristics of the closed-loop AI inertial sensor for a 40000-s portion of night data. The Allan deviation of the rotation rate is shown as the blue line. The Allan deviation of the acceleration before temperature compensation is shown as the black line, and the red line shows the Allan deviation of temperature compensated signals.}		
	\end{figure}
 
	For the directly measured inertial sensor based on a dual AI, the interference phases of the two atomic beams are calculated according to the signal $S_1$ and $S_2$ and the fitted $A$ and $B$. The changes in rotation rate and acceleration can be deduced by adding or subtracting the two phases. For the inertial sensor operating in acceleration-closed-loop mode, the acceleration can be obtained via the output signal from the PID controller, and the rotation rate needs to be calculated by solving the phase of $S_1-S_2$ according to Eq. (\ref{equ:sum and different singal}). Short-term sensitivities are determined by operating the AI inertial sensor in open-loop mode and first calculating the Allan standard deviations of the two AIs, The average short-term sensitivities are calculated to be $\sim 0.56 \ \rm{mrad/s/\sqrt{Hz}}$ for the two open-loop AI gyroscopes and $\sim1.2$ m$g$ for the two open-loop AI accelerometers.
 
    To evaluate the performance of the closed-loop dual-AI inertial sensor, we calculate the Allan standard deviations of the rotation rate and the acceleration. Over a 40000-s acquisition time, the Allan standard deviation of the rotation rate (blue line) reaches a short-term sensitivity of 62 $\rm{\mu rad/s/\sqrt{Hz}}$ and demonstrates a long-term stability of approximately 840 nrad/s, as shown in Fig. \ref{fig:rotation rate}. This result shows better performances than with the open-loop method. This is because the dual-AI gyroscope operating with acceleration-phase-locked approach will compensate for acceleration phase drifts in real time, which can dynamically compensate for noise introduced by mirror vibration caused by environmental vibration, mirror creep, and other factors. The Allan standard deviations for the closed-loop AI accelerometer (black lines) are also shown in Fig. \ref{fig:rotation rate}, and are directly calculated from the output signal of the PID controller. The sum signal is feedback for closed-loop control, so the noise from the dual AIs are superimposed. This approach makes the accelerometer output more sensitive than the gyroscope to noise caused by changes in environmental factors such as temperature. Therefore, the short-term sensitivity of the accelerometer is only $0.39 \ \rm{m}$$g$$\rm{/\sqrt{Hz}}$, and the long-term stability is $0.78 \ \rm{m}$$g$. To minimize the influence of temperature variations, we strategically positioned four temperature sensors: three at different locations surrounding the vacuum system and one outside the sensor head to measure ambient temperature. Temperature correction for inertial sensor effects was achieved through training a random forest model during an initial calibration period. Upon completion of this calibration phase, we conducted temperature prediction and compensation during subsequent data collection, maintaining constant temperature calibration coefficients. This approach was implemented to consistently alleviate the impact of temperature changes on the inertial sensor readings. The Allan standard deviation of the accelerometer after compensation is shown in Fig. \ref{fig:rotation rate}(red line), with a short-term sensitivity of $30 \ \rm{\mu}$$g$$\rm{/\sqrt{Hz}}$ and long-term stability of $6.1 \ \mu g$.

	\begin{figure}[t]
		\includegraphics[width=\linewidth]{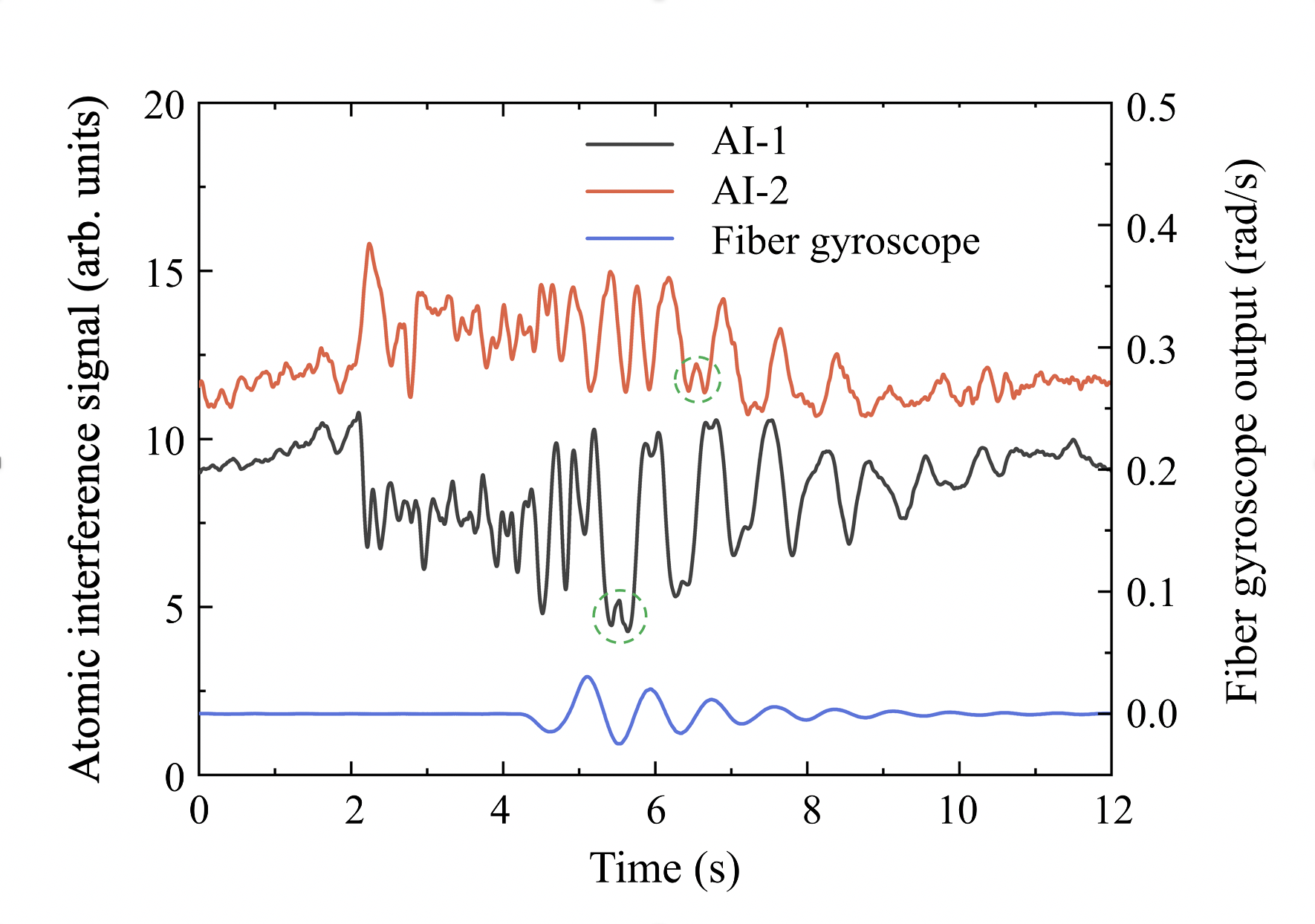}
		\caption{\label{fig:rotation} Measurement of the dynamic rotation rates. The signals of the AI (black and red) and the fiber-optic gyroscope (blue) show oscillation when a force is applied to the platform. Ambiguity of the interferometric sensor appears in the peaks and troughs (green dashed circles).}
	\end{figure}

\begin{figure}[t]
		\includegraphics[width=\linewidth]{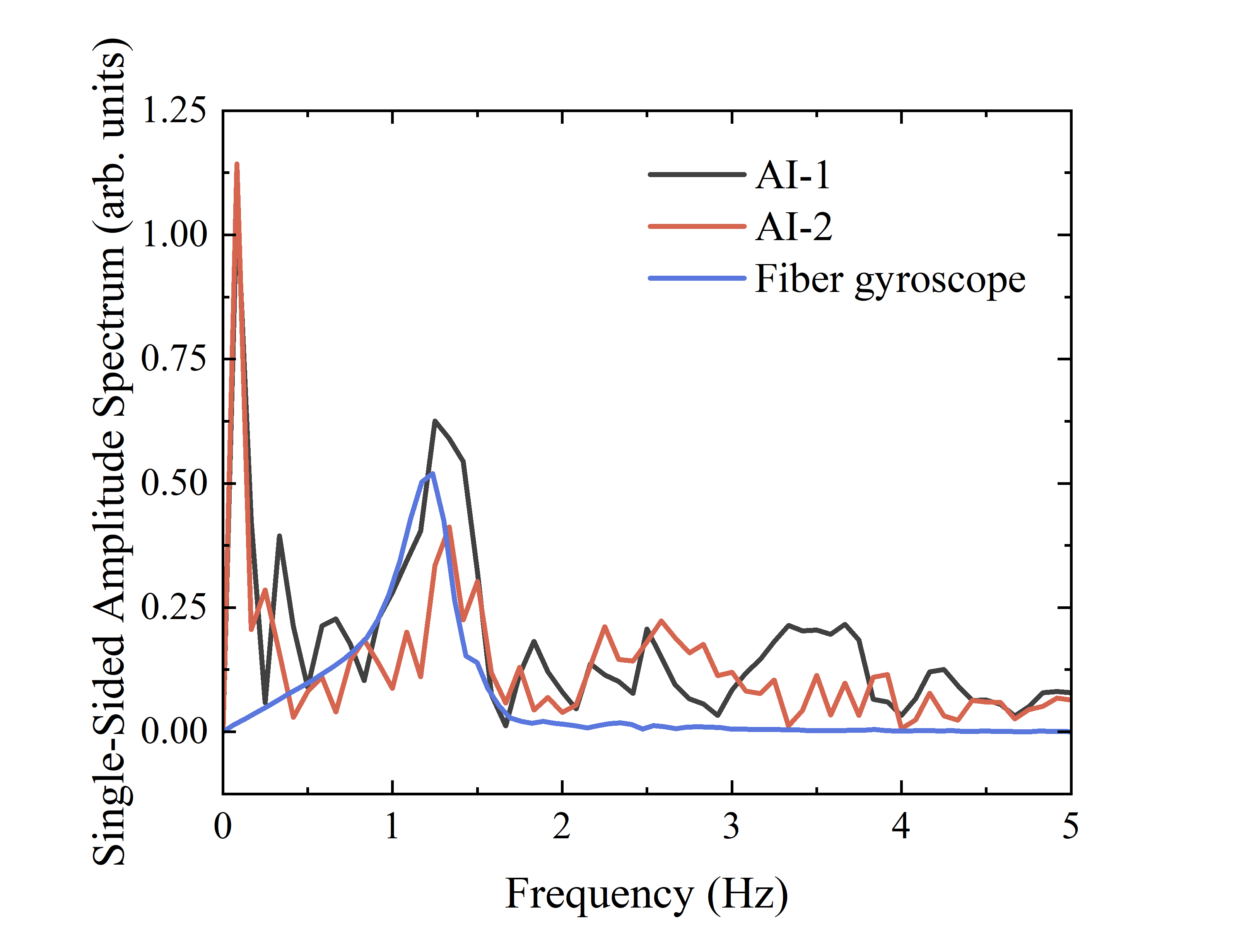}
		\caption{\label{fig:rotation_fft} Fourier analysis of the total rotation rate signal of Fig. \ref{fig:rotation}, with a frequency resolution of 83 mHz for the dual AIs and 65 mHz for the fiber-optic gyroscope.}
	\end{figure}

	A rotational test was performed to verify the inertial sensitivity of the AI gyroscope. The entire sensor was placed on a platform, which was floating on four pneumatic vibration isolators (I-2000, Newport, USA) and adjusted to be perpendicular to the direction of local gravitational acceleration using a tilt sensor. A fiber-optic gyroscope was fixed on the same platform to measure the rotation rate simultaneously using the same sensing axis as the AI gyroscope. The AI gyroscope was operated in open-loop mode. Figure \ref{fig:rotation} shows the signals measured with the AI gyroscope and the fiber-optic gyroscope when an external force was applied to turn the entire platform around the central axis. The two AIs pick up the change in the rotation rate induced by the applied force and output two oscillating signals with frequencies of 1.25 Hz and 1.33 Hz deduced from their Fourier transform spectra, as shown in Fig. \ref{fig:rotation_fft}. The output signal from the fiber-optic gyroscope also responds to the applied force with an oscillation frequency of 1.33 Hz, which agrees well with the AI results. In addition, some umbilications appear in the peaks and troughs of the dual-AI signals that are caused by the measurement ambiguity of the interferometric sensor, as shown in Fig. \ref{fig:rotation} (green dashed circles). Therefore, the atomic interference signal in Fig. \ref{fig:rotation_fft} not only includes the fundamental frequency, but also the second harmonic signal. When the output of the fiber-optic gyroscope is compared with that of the AI gyroscopes, the open-loop dynamic range of the AI is estimated to be $\sim$0.0157 rad/s, which is slightly different from the theoretical value (0.0118 rad/s). This may be due to the difference between the theoretical and measured values of the AI scale factor, such as the measurement error in atomic velocity. In addition, the sensing axes of the two gyroscopes may be not completely consistent, and the measurement may not be synchronized completely.
 
	Potential further improvements in performance of the AI inertial sensors are mainly limited by the short interrogation time ($T=0.87$ ms) and the correspondingly small interference area ($A=0.097$ mm$^2$). A larger interference area will increase the scale factor of each AI inertial sensor and improve its sensitivity. However, the contrast of the interference fringes may decrease with increasing interrogation time because of the expansion of the atomic beams, which leads to a deterioration in sensitivity. The sensitivity can be improved by increasing the number of atoms involved in the Raman transition, which is affected by the Raman laser beam waist because of Raman velocity selection. In our case, the Doppler-sensitive Raman transition linewidth (FWHM) for the $\pi$ pulse is measured to be 13.7 kHz with a Raman beam waist of 1 mm, and the deduced selected horizontal velocity linewidth (FWHM) is approximately 0.54 cm/s. This means that only a small percentage of the atoms contribute to the interference signal, considering the horizontal velocity distribution (FWHM) of 8.7(2) cm/s for our cold atomic beams. In future work, the percentage of atoms involved in the interference and thus the signal contrast can be increased both by reducing the beam waist size of the Raman lasers and narrowing the horizontal velocity distributions of the atomic beams. 
	
	\section{Conclusion}
	
	In this work, we demonstrated a dual-AI inertial sensor with closed-loop phase-locking that uses continuous cold atomic beams. The differential mode with dual AIs suppresses common-mode noise  effectively and enhances the sensitivity and stability of the gyroscope in combination with closed-loop acceleration-phase-locking method. The short-term sensitivity and long-term stability of the accelerometer have also been improved through temperature monitoring and compensation. The acceleration and the rotation rate can be simultaneously decoupled and measured by locking the differential signal of the dual AIs, and reach long-term stabilities of 840 nrad/s for the gyroscope and 6.1 $\mu g$ for the accelerometer with an interrogation time of only 0.87 ms. Continuous-operation mode helps in achieving a high-bandwidth zero-dead-time AI inertial sensor. Using the AI inertial sensor, we measured the rotation signal continuously when the platform was shaking at a frequency of approximately 1.33 Hz, with results that were consistent with the output from the fiber-optic gyroscope.
	
	Closed-loop measurement of the acceleration based on continuous AI inertial sensors was demonstrated, while the measurement of the rotation rate was measured in an open-loop mode. Controlling the phase of the radio-frequency signal that drives the EOM can realize closed-loop control of the initial phase of the Raman laser and real-time feedback of the rotation phase, thus enabling double closed-loop decoupled measurements of the acceleration and the rotation rate. In future work, using a narrower Raman beam or a three-dimensional sub-Doppler-cooled atomic beam may enable higher fringe contrast, thereby leading to higher sensitivity. In addition, the sensor's measurement bandwidth can be increased by increasing the modulation frequency, which can then improve the response speed of the AI inertial sensor for use in a dynamic environment. This closed-loop dual-AI approach represents a promising way to realize an AI inertial measurement unit for use in field applications that require a large dynamic range, high accuracy and high stability, providing a basic element that produces simultaneous decoupled outputs of the acceleration and the rotation rate.
	
	\begin{acknowledgments}
		This  work  was  supported  by  the  National  Natural  Science Foundation of China (Grant No.61473166).
	\end{acknowledgments}
	
	\appendix
	



\end{document}